\documentclass[conference]{IEEEtran}
\IEEEoverridecommandlockouts
\usepackage{cite}

\usepackage{booktabs} 
\usepackage{array} 
\usepackage{multirow}
\usepackage{amsmath,amssymb,amsfonts}
\usepackage{algorithmic}
\usepackage{graphicx}
\usepackage{textcomp}
\usepackage{xcolor}
\usepackage{url}
\usepackage[hidelinks]{hyperref}

\usepackage{pifont}

\def\BibTeX{{\rm B\kern-.05em{\sc i\kern-.025em b}\kern-.08em
    T\kern-.1667em\lower.7ex\hbox{E}\kern-.125emX}}
\begin{document}

\title{Open-Source 5G RAN Platforms: A Dual Perspective on Performance and Capabilities}

\author{\IEEEauthorblockN{1\textsuperscript{st} Maria Barbosa}
\IEEEauthorblockA{\textit{Centro de Informática} \\
\textit{Universidade Federal} \\ \textit{de Pernambuco} (UFPE)\\
Recife, Brasil \\
mksb@cin.ufpe.br}
\and
\IEEEauthorblockN{2\textsuperscript{nd} Iasmin Gomes}
\IEEEauthorblockA{\textit{Centro de Informática} \\
\textit{Universidade Federal} \\ \textit{de Pernambuco} (UFPE)\\
Recife, Brasil \\
imgs@cin.ufpe.br}
\and
\IEEEauthorblockN{3\textsuperscript{rd} Vinícius Melo}
\IEEEauthorblockA{\textit{Centro de Informática} \\
\textit{Universidade Federal} \\ \textit{de Pernambuco} (UFPE)\\
Recife, Brasil \\
vlsm@cin.ufpe.br}
\and
\IEEEauthorblockN{4\textsuperscript{th}  Kelvin Dias}
\IEEEauthorblockA{\textit{Centro de Informática} \\
\textit{Universidade Federal} \\ \textit{de Pernambuco} (UFPE)\\
Recife, Brasil \\
kld@cin.ufpe.br}
}

\maketitle

\begin{abstract}
Open-source implementations of the fifth generation (5G) Radio Access Network (RAN), such as OpenAirInterface (OAI) and srsRAN, have become increasingly relevant for industry and academic research, rapid prototyping, and low-cost private 5G deployments. The state of the art presents different evaluation scenarios for both platforms; however, these are often limited to individual testbed evaluations, end-to-end studies, or partial comparisons between them. In particular, there is a lack of evaluations considering different Software-Defined Radios (SDRs), RAN configurations, user applications, and numbers of real UEs. To address this gap, this paper presents a comparative evaluation of OAI and srsRAN. The study includes a qualitative assessment of supported features and deployment flexibility, followed by a quantitative performance analysis covering radio resource control setup procedures, compliance with theoretical data rates, and real application workloads with diverse quality of service requirements. Specifically, we consider Video on Demand (VoD) for data rate evaluation, Live Streaming (LS) and Cloud Gaming (CG) as latency-sensitive applications. These workloads provide a comprehensive view of each platform capability.
\end{abstract}


\section{Introduction}
\label{sec:intro}


The Radio Access Network (RAN) in the fifth generation of cellular networks (5G) is composed of the Next-Generation Node B (gNodeB), which provides connectivity between the User Equipment (UE), and the 5G Core (5GC). The gNodeB consists of a base station equipped with 5G air interface, known as New Radio (NR). 3rd Generation Partnership Project (3GPP) developed NR to meet the minimum requirements of 5G services categories defined by the International Telecommunication Union (ITU). In 3GPP Release 15, the first for 5G, the focus was on Enhanced Mobile Broadband (eMBB) services, aiming to improve data rates for end-users, such as in Video on Demand (VoD) applications. Release 16 extended the scope to Ultra Reliable Low Latency Communications (URLLC), targeting applications such as live streaming (LS), and cloud gaming (CG), as well as Massive Machine-Type Communications (mMTC) services \cite{release15}.

To meet these requirements, NR operates in two frequency ranges: FR1, also known as sub-6 GHz, which includes low-band (below 1 GHz) and mid-band (1–6 GHz), and FR2, above 24 GHz, referred to as millimeter waves band. Furthermore, in Long Term Evolution (LTE) Release 14, 3GPP introduced the functional split in the RAN, dividing the protocol stack into the Central Unit (CU), Distributed Unit (DU), and Radio Unit (RU). Until then, the RAN was based on a monolithic Baseband Unit (BBU), highly dependent on proprietary solutions. With this paradigm shift and the standardized interfaces defined by 3GPP, it became possible to deploy virtualized environments using Commercial Off-The-Shelf (COTS) hardware. In this context, open-source platforms such as srsRAN \cite{srs} and OpenAirInterface (OAI) \cite{oai} emerged. These platforms, combined with Software-Defined Radios (SDRs), such as Ettus Research Universal Software Radio Peripheral (USRPs) B210 and N310, provide the framework for implementing low-cost private 5G networks, widely applicable in both industrial and academic \cite{core_eval} environments.

Both 5G RAN platforms comply with 3GPP specifications but provide different features and remain under continuous development. Therefore, it is essential to analyze them to understand their respective benefits and limitations. In the literature, several works evaluate the performance of these platforms individually in different scenarios. For instance, \cite{eval_testbed} and \cite{bench} compare real and simulated implementations of OAI-based testbeds. In \cite{perform_testbed}, different standalone (SA) 5G testbeds based on OAI are proposed, analyzing performance concerning USRP interoperability, RAN configurations, and UE types. Other studies focus on the interoperability between open-source RAN and 5GC combinations, such as \cite{comp_e2e} and \cite{oss}, which also provide a qualitative assessment of OAI and srsRAN, including compliance with Open RAN. In \cite{comp_ran} presents a comparative analysis of two SDR-based open-source platforms for 5G testbeds, exploring their integration with virtualized environments, particularly Docker\cite{docker}. Finally, \cite{comp_ran} provides a comparative analysis of the two platforms employing the USRP B210 under two distinct bandwidth configurations, namely 20 MHz and 40 MHz, evaluated through iPerf \cite{iperf3}, ping, and Google Meet call scenarios. Nonetheless, it does not address MIMO configuration capabilities, and all experiments are conducted with a single UE.

Despite these contributions, there is still a gap in the state of the art regarding the evaluation of such platforms using real applications, such as VoD, LS, and CG. Moreover, few studies assess RAN control procedures, such as Radio Resource Control (RRC) setup, across different USRPs and multiple real UEs. To address this gap, this article aims to provide a qualitative analysis of the leading open-source 5G RAN platforms, highlighting their features, capabilities, and limitations. Next, we propose a low-cost standalone 5G network prototype, using a COTS server to run the 5GC and open-source RAN platforms together with USRP B210 and N310 radios, under different radio configurations such as Multiple Input Multiple Output (MIMO) and varying bandwidths. The prototype delivers VoD, LS, CG, and iPerf3 \cite{iperf3} services to the users. Finally, we perform a quantitative evaluation, including RRC setup in the control plane and multiple assessments in the data plane, analyzing how each platform, under the proposed configurations, approaches the theoretical values defined by 3GPP and behaves with different USRPs and real applications.

The next sections are structured as follows: Section \ref{sec:Qualitative} details the 5G RAN platforms. The conception of a low-cost prototype for the 5G network is presented in Section \ref{sec:rede5G}. Section \ref{sec:Evaluation} presents the performance evaluation results. Finally, the conclusions and future work are presented in Section \ref{sec:conc}.
\section{Open Source 5G RAN Platforms}
\label{sec:Qualitative}

This section provides an overview of open-source 5G RAN platforms. Table \ref{tab:analise_qualitativa} highlights the main characteristics of each platform, divided into two groups. The first outlines general specifications, including 3GPP release, installation difficulty, configuration difficulty, programming language used for development, and virtualization infrastructure (VI). The second group presents currently supported features, such as Sub-Carrier Spacing (SCS), MIMO, 3GPP functional split 7.2, Data Plane Development Kit (DPDK), Time and Frequency Division Duplex (TDD/FDD), Network Slicing, and preliminary Non-Terrestrial Networks (NTN) support. 

\begin{table}[h!]
\centering
\caption{Comparison of Open-Source 5G RAN Platforms.}
\begin{tabular}{c c c}
\hline
\textbf{Parameter} & \textbf{OAI} & \textbf{srsRAN} \\
\hline
\textbf{3GPP Release} & Rel-15/16/17 & Rel-17 \\
\textbf{Installation Difficulty} & Medium & Low \\
\textbf{Configuration Difficulty} & High & Low  \\
\textbf{Language} & C & C++ \\
\multirow{2}{*}{\textbf{Virtualization Infrastructure}} & Container & Container\\
& and Pod & \\
\hline
\multicolumn{3}{c}{\textbf{Supported Features}} \\
\hline
\multirow{2}{*}{\textbf{SCS}} & 15/30 kHz - FR1 & 15/30 kHz - FR1 \\
& 120 kHz - FR2 & \\
\textbf{MIMO} & up to 2x2 & up to 4x4 \\
\textbf{3GPP Split 7.2} & \checkmark & \checkmark \\
\textbf{DPDK} &  \checkmark &  \checkmark \\
\textbf{TDD/FDD} & \checkmark &  \checkmark  \\
\textbf{QAM-256} & \checkmark &  \checkmark  \\
\textbf{Network Slicing} & \checkmark &  \checkmark  \\
\textbf{NTN (preliminary)} & \checkmark &  \checkmark  \\
\hline
\end{tabular}
\label{tab:analise_qualitativa}
\end{table}

The OpenAirInterface (OAI) Software Alliance provides both RAN and 5GC implementations. The RAN is an open-source project written in C, supporting bandwidths up to 100~MHz, TDD/FDD operation, 256-QAM modulation, and 2x2 MIMO. One of its distinguishing features is experimental support for 120~kHz SCS in FR2. Additionally, OAI offers flexible deployment options, including bare-metal installation, Docker \cite{docker} containerization, and Kubernetes pods. Regarding installation, OAI provides pre-configured Dockerfiles and required build flags, keeping the installation difficulty at a moderate level. As for configuration, there are several configuration files available, however, comprehending the detailed functionality of each parameter remains challenging. Nevertheless, the development documentation is extensive and well-detailed, covering compliance aspects and implementation guidelines.

The srsRAN project, developed by Software Radio Systems (SRS), is an open-source 5G CU/DU that includes the full L1/L2/L3 stack. It supports both FDD and TDD across FR1 bands and offers bandwidths up to 100~MHz (e.g., 100~MHz TDD, 50~MHz FDD). It is compliant with selected features of 3GPP Release~17 and is implemented in C++. Installation and configuration are generally simpler compared to OAI. The official guide is concise and provides two installation options: building from source or using Ubuntu packages. Although srsRAN does not provide many configuration examples, it includes a comprehensive configuration reference, offering an overview of available parameters. Among its notable supported features are 3GPP split~7.2 and up to 4x4 MIMO.  

\section{5G Network Prototype}
\label{sec:rede5G}

\begin{figure}[ht!]
    \centering
    \includegraphics[width=0.95\linewidth]{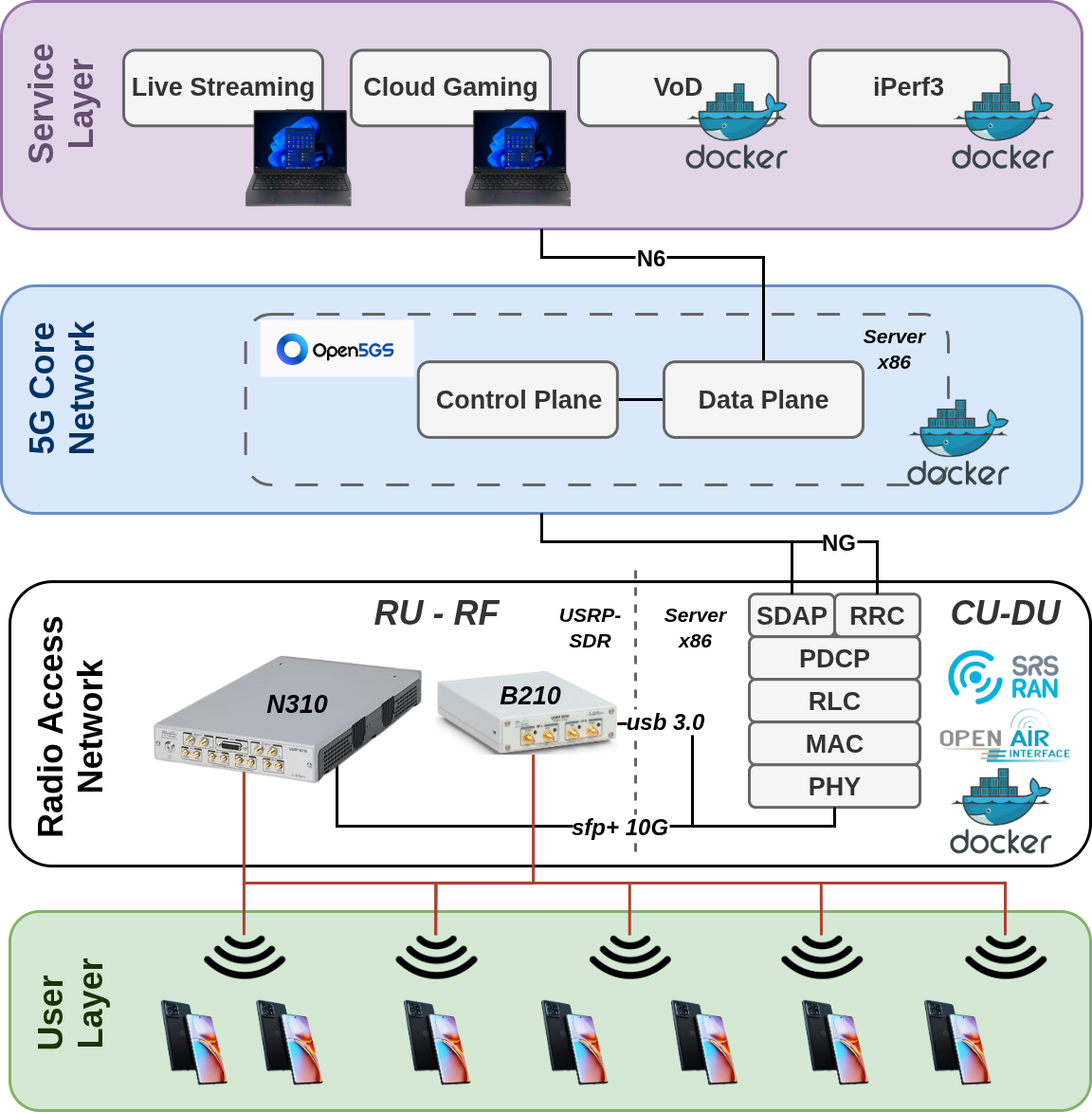}
    \caption{Architecture of the 5G SA prototype network.}
    \label{fig:env}
\end{figure}

This section presents the hardware and software components used to design and deploy a 5G SA prototype network to evaluate the performance of the RAN with two open-source platforms: OAI and srsRAN. Figure \ref{fig:env} illustrates the proposed testbed architecture, organized into four layers: service, 5GC, RAN, and the user layer composed of mobile devices (UEs). 

The service layer provides different applications to the users, enabling performance evaluation under diverse workloads, from simple tools such as iPerf3 \cite{iperf3} to more demanding applications such as VoD, LS, and CG. The 5GC handles control-plane procedures (authentication, registration, and Packet Data Unit (PDU), session establishment) as well as data-plane functionalities, including interconnection with the Data Network (DN) through the N6 interface. 

The RAN follows the 3GPP functional split Option 8, where the CU and DU implement the radio protocol layers: Physical (PHY), Medium Access Control (MAC), Radio Link Control (RLC), Packet Data Convergence Protocol (PDCP), Radio Resource Control (RRC), and Service Data Adaptation Protocol (SDAP). This part of the RAN was deployed with OAI and srsRAN in a Docker-based \cite{docker} environment on an x86 server. The RU, responsible only for radio-frequency (RF) functions, was implemented with two SDRs: the USRP Ettus B210 and N310. The SDRs convert baseband signals processed on the server into RF signals (and vice versa), enabling 5G communication with real UEs through antennas. 

Table \ref{tab:confenv} summarizes the testbed specifications, including the Virtual Infrastructure (VI), software versions, and the mobile device. The deployment leverages Docker \cite{docker} network to interconnect the 5GC and the gNodeB.  

\begin{table}[h!]
    \centering
        \caption{Specifications of the 5G testbed configuration.}
    \begin{tabular}{c p{3cm} c}
    \hline
     & \textbf{Component} &  \textbf{Specification} \\
    \hline
     \multirow{5}{4em}{\textbf{VI}} & CPU & Intel Xeon Gold 5215 \\ 
    & RAM & 96 GB \\ 
    & Docker & 27.5.1 \\
    & Docker Compose & 2.27.0 \\
    & Operating System & Ubuntu 22.04.5 LTS \\
    \hline
     \multirow{2}{4em}{\textbf{5GC}} & Platform & Open5GS \\ 
    & Version & v2.7.2 \\
    \hline
    \multirow{2}{4em}{\textbf{RAN}} & Antenna & ANT-5GWWS3-SMA \\
    & UHD Version & 4.1.0.0\\
    \hline
    \multirow{2}{4em}{\textbf{User}} & Model & Motorola Edge 20 \\
    & SIM Card & Sysmocom S1J1 \\
    \hline
    \end{tabular}
    \label{tab:confenv}
\end{table}

\subsection{Radio Configuration}

Radio configuration directly impacts network performance. According to 3GPP Technical Specification (TS) 38.306, the achievable single-carrier data rate ($R$) is given in \cite{3gpp_38_306}. As presented in the TS, $R$ mainly depends on the modulation, the number of Physical Resource Blocks (PRBs), and the number of layers ($\nu_{L}$). In this work, we adopted a fixed modulation of 256-QAM and an SCS of 30~kHz, testing three configurations: a single layer with bandwidths of 20 and 40~MHz, and 2$\times$2 MIMO with 20~MHz. Note that 2$\times$2 MIMO was not combined with the 40~MHz bandwidth, since operating in this mode requires a 46.08~MHz master clock, which exceeds the 30.72~MHz maximum supported by the USRP B210 with two TX channels. These configurations provide a broader view of the performance of each stack under different operational settings and resource allocation conditions.

In 5G, typically, mid-bands such as n78 (3.3–3.8~GHz) adopt TDD. In TDD, uplink (UL) and downlink (DL) share the same frequency, with slot-based separation in time, forming a TDD pattern. This pattern defines the slot allocation for DL and UL, which directly affects the achievable data rates. The TDD pattern used in this work is [D D D D D D D F U U]. The symbol F denotes a flexible slot, which can be assigned to either DL or UL. In our configuration, the flexible slot was structured to allocate eight symbols to the DL and four to the UL. Finally, Table \ref{tab:confs} presents the combination of the configurations.

\begin{table}[h!]
    \centering
    \caption{Configurations summary.}
    \begin{tabular}{c c c c c}
    \hline
        \textbf{Configuration Number} & \textbf{MCS} & \textbf{SCS} & \textbf{Bandwidth} & $\nu_{L}$ \\
        \hline
        Configuration 1 (C1) & 256 & 30 kHz & 20MHz & 1 \\
        \hline
        Configuration 2 (C2) & 256 & 30 kHz & 40MHz & 1 \\
        \hline
        Configuration 3 (C3) & 256 & 30 kHz & 20MHz & 2 \\
    \hline
    \end{tabular}
    \label{tab:confs}
\end{table}

\section{Performance Evaluation} 
\label{sec:Evaluation}

This section details the methodology for evaluating the performance of the RAN platforms considered in this article, followed by the results and comparative analysis. The evaluation covers a control plane metric, time for RRC setup procedure, and data plane metrics, divided into two main groups, the theoretical flow to verify how close the stack is of the theoretic data rate defined by the 3GPP for throughput and, and performance under real application workloads, such as VoD, LS and CG.

\subsection{Case 1: RRC Setup procedure}
The first part of the performance evaluation consists of analyzing the time required to complete the RRC Setup procedure, whose objective is to identify which combination achieves the lowest average value. The procedure begins when the UE sends the \textit{RRCSetupRequest} message through the $UL\_CCCH$ channel. In response, the gNodeB checks the available configurations for the UE and transmits the \textit{RRCSetup} message through the $DL\_CCCH$, containing parameters such as modulation, bandwidth, and TDD pattern. Then, the UE processes this information and completes the procedure by sending the \textit{RRCSetupComplete} message, after which Non-Access Stratum (NAS) messages can be exchanged with the 5GC.

\begin{figure}[h!]
    \centering
    \includegraphics[width=0.9\linewidth]{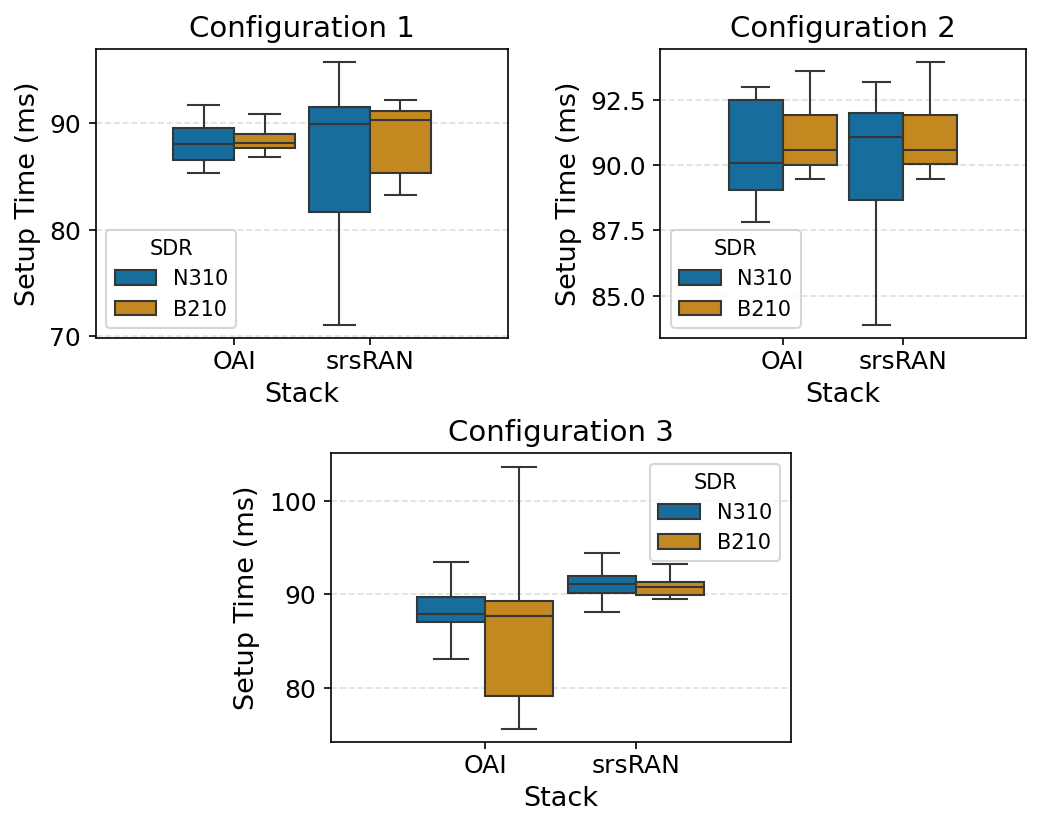}
    \caption{Time of the RRC Setup procedures.}
    \label{fig:rrc_setup}
\end{figure}

The experiment involved four different UEs, each subjected to ten connection and disconnection cycles. To ensure consistency, a Python script automated the activation and deactivation of airplane mode, allowing the four devices to connect and disconnect simultaneously and enabling the evaluation of resource competition. The results are presented in Figure \ref{fig:rrc_setup}.

The figure shows that, on average, configuration C1 achieved the best performance, with 87.7 ms, both in srsRAN and OAI. Next, C3 presented 88.5 ms, and C2, 91.9 ms. The best isolated result occurred with configuration C3 using the B210 board in the OAI platform, registering 85.3 ms. The second-best result, obtained by C1 with srsRAN on the N310, showed an increase of only 1.65\%, being statistically equivalent to the first. On the other hand, the worst performance occurred in configuration C2 with OAI on the N310, whose average time was 15.2\% higher than the best result. Overall, the configurations exhibited similar behavior for this metric.

\subsection{Case 2: Theoretical Throughput}
The second part of the performance evaluation focuses on the theoretical throughput (TT), defined as as the maximum achievable data rate for a given NR configuration under ideal conditions; values were computed following 3GPP TS~38.306 \cite{3gpp_38_306}.
The experiment used iPerf3 \cite{iperf3} to measure the end-to-end capacity between the UE and the server, presented in Section \ref{sec:rede5G}. For each combination, we tested both downlink and uplink flows. The target data rate in each test was set according to the TT expected and the test duration was fixed at 100 seconds with a reporting interval of 1 second. The results, expressed in terms of average throughput $\bar{x}$, standard deviation $s$, throughput efficiency given by the ratio between the achieved and the theoretical throughput $\% = \frac{\bar{x}}{TT}$, and spectral efficiency (SE) defined as $\frac{\bar{x}}{BW}$, expressed in bits/s/Hz, are summarized results are presented in Table \ref{tab:iperf_results}.

\begin{table}[htb!]
\centering
\caption{iPerf Results with Spectral Efficiency.}
\label{tab:iperf_results}
\begin{tabular}{ccccccccccc}
\toprule
\multicolumn{11}{c}{\textbf{Downlink}} \\
\midrule
\textbf{SDR} & \textbf{Conf.} & \textbf{TT}
& \multicolumn{4}{c}{\textbf{OAI}}
& \multicolumn{4}{c}{\textbf{srsRAN}} \\
\cmidrule(lr){4-7} \cmidrule(lr){8-11}
& & & $\bar{x}$ & SE & $s$ & \% & $\bar{x}$ & SE & $s$ & \% \\
\midrule
\multirow{3}{*}{B210}

& 1 & 84  & 51.7 & 2.59 & 7.786 & 61 & 61.45 & 3.1 & 5.437 & \textbf{73} \\
& 2 & 175 & 118 & 2.95 & 10.45 & 67 & 126.9  & 3.2 & 12.47 & \textbf{72} \\
& 3 & 168 & 55.7 & 2.79 & 6.472 & 33 & 121.8 & 6.1 & 10.53 & \textbf{72} \\
\addlinespace
\multirow{3}{*}{N310}
& 1 & 84  & 49.42 & 2.47 & 6.912 & 58 & 61.43 & 3.1 & 3.888 & \textbf{73} \\
& 2 & 175 & 137.0 & 3.43 & 15.50 & \textbf{78} & 122.8 & 3.1 & 19.18 & 70 \\
& 3 & 168 & 53.35 & 2.67 & 10.38 & 31 & 121.8 & 6.1 & 9.271 & \textbf{72} \\

\midrule
\multicolumn{11}{c}{\textbf{Uplink}} \\
\midrule
\textbf{SDR} & \textbf{Conf.} & \textbf{TT}
& \multicolumn{4}{c}{\textbf{OAI}}
& \multicolumn{4}{c}{\textbf{srsRAN}} \\
\cmidrule(lr){4-7} \cmidrule(lr){8-11}
& & & $\bar{x}$ & SE & $s$ & \% & $\bar{x}$ & SE & $s$ & \% \\
\midrule
\multirow{3}{*}{B210}
& 1 & 25 & 9.251 & 0.46 & 2.476 & 37 & 10.24 & 0.51 & 2.508 & \textbf{40} \\
& 2 & 52 & 14.36 & 0.36 & 2.139 & 27 & 16.29 & 0.41 & 5.212 & \textbf{31} \\
& 3 & 50 & 10.09 & 0.50 & 3.718 & \textbf{20} & -- & -- & -- \\

\addlinespace
\multirow{3}{*}{N310}
& 1 & 25 & 2.611 & 0.13 & 1.614 & 10 & 8.029 & 0.40 & 3.818 & \textbf{32} \\
& 2 & 52 & 12.90 & 0.32 & 3.366 & 24 & 25.95 & 0.65 & 3.611 & \textbf{49} \\
& 3 & 50 & 2.805 & 0.14 & 1.808 & 5 & 11.82 & 0.59 & 3.335 & \textbf{23} \\

\bottomrule
\end{tabular}
\end{table}

The results analysis reveals significant performance disparities between the evaluated stacks, with srsRAN demonstrating superior consistency across most scenarios. In C1, srsRAN achieved downlink throughput values 18-24\% higher than OAI for B210 and N310, respectively. However, in C2, OAI demonstrated superior performance, surpassing srsRAN by 12\% in N310 downlink while maintaining competitive results in B210 with only 7\% lower throughput. The most pronounced difference occurred in C3, with OAI achieving only 45.7\% and 43.8\% of srsRAN throughput for B210 and N310, respectively. 

OAI exhibited a performance drop in C3, mainly due to instability issues encountered during the MIMO tests, which were also evident in other experiments presented in this paper. Regarding the individual performance of each platform, it can be observed that, in terms of proximity to the theoretical throughput, srsRAN achieved results closer to the theoretical values in the downlink scenarios, reaching approximately 70\% of the expected throughput.

Regarding uplink performance, all scenarios showed limitations in all scenarios, with throughput values reaching only 5-49\% of theoretical capacity, indicating substantial constraints in the uplink transmission capabilities of the evaluated 5G implementations. Nevertheless, srsRAN maintained a more consistent performance across configurations, with the exception of the scenario using the B210 with configuration c3, which exhibited severe instability and could not be completed.

Finally, regarding SE, it was observed that srsRAN achieved the best results in the downlink, as expected, given its superior throughput performance. Interestingly, srsRAN also maintained its efficiency even when switching between different SDR boards, indicating that the main impact factor lies within the RAN configuration parameters rather than the hardware itself.

\subsection{Case 3: Performance under Real Application Workloads}
This subsection presents the results of the performance of the RAN platforms under different real applications workloads. The workloads were divided in three application, each one with different QoS requirements, such as throughput and latency. The goal of this evaluation is to identify the most suitable combination to provide the requirements of each application.   

\subsubsection{Video on Demand (VoD)}

\begin{figure*}[ht!]
    \centering
    \includegraphics[width=0.85\linewidth]{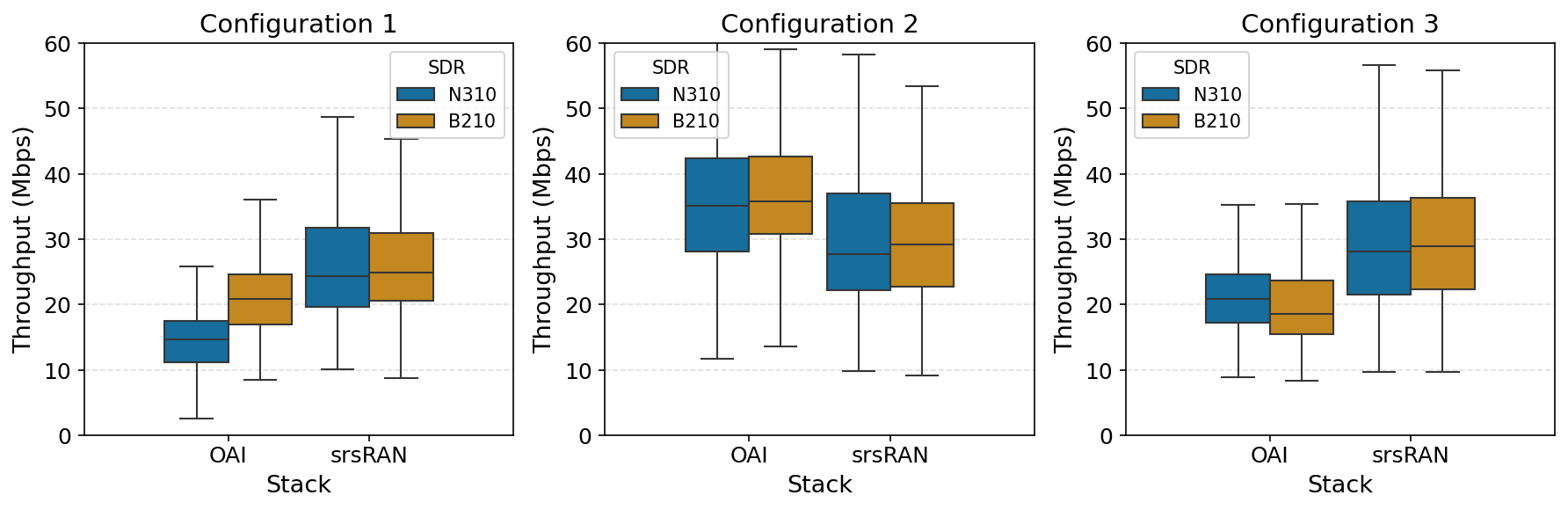}
    \caption{Throughput comparison for VoD across different configurations and 5G stacks.}
    \label{fig:vod_data_rate}
\end{figure*}

\begin{figure*}[ht!]
    \centering
    \includegraphics[width=0.85\linewidth]{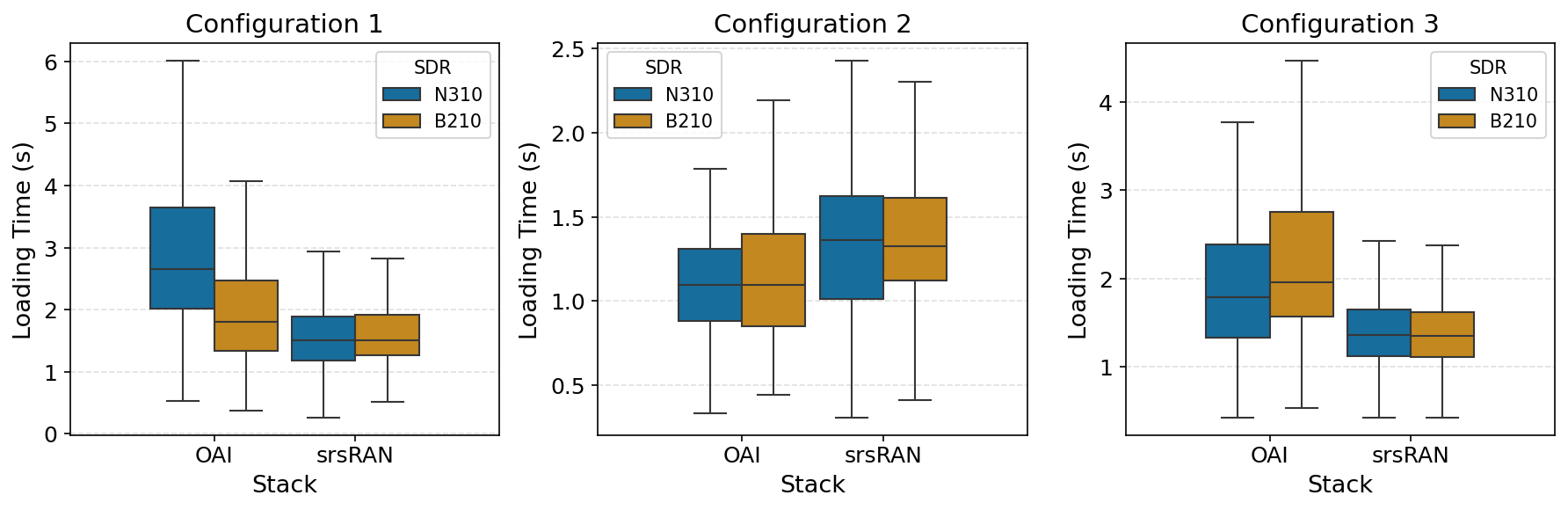}
    \caption{Loading time comparison for VoD across different configurations and 5G stacks.}
    \label{fig:vod_load_time}
\end{figure*}

VoD consists of sending video content to the user in segmented chunks. Each chunk, when sent, requires high throughput and a certain loading time to be downloaded. The chunks are played on the video screen, and when the current playing is about to end, a new chunk is requested. The VoD Server is a Flask\footnote{https://flask.palletsprojects.com/en/3.0.x/}-based Python application that provides the video and records QoS metrics. On the user side, a Java-based Android application requests a video (Big Buck Bunny 60fps/1080p) from the server, downloads it, and watches it on the phone. The application calculates the network metrics and then sends it to the server. Finally, to create a competitive environment, four UEs were used simultaneously request video chunks from the server. Each UE downloaded about 60 chunks, each lasting 10 seconds.

Figures \ref{fig:vod_data_rate} and \ref{fig:vod_load_time} present the throughput and loading time results for video chunks download, respectively. In C1, srsRAN exhibited superior performance over OAI, particularly with N310. While OAI required more than 2.8 seconds for 50\% of video chunks to load, srsRAN achieved more stable performance with median loading times around 1.5 second. In contrast, C2 provided the most balanced performance across both stacks and SDRs, demonstrating higher throughput and lower loading times compared to other configurations. Finally, C3 produced mixed outcomes, with srsRAN maintaining better loading time consistency across both SDRs, whereas OAI showed increased variability. Overall, the competitive environment with four simultaneous UEs revealed that srsRAN maintained better QoS stability under load, while OAI performance was more sensitive to configuration parameters.

\subsubsection{Live Streaming}

\begin{figure}[ht!]
    \centering
    \includegraphics[width=0.85\linewidth]{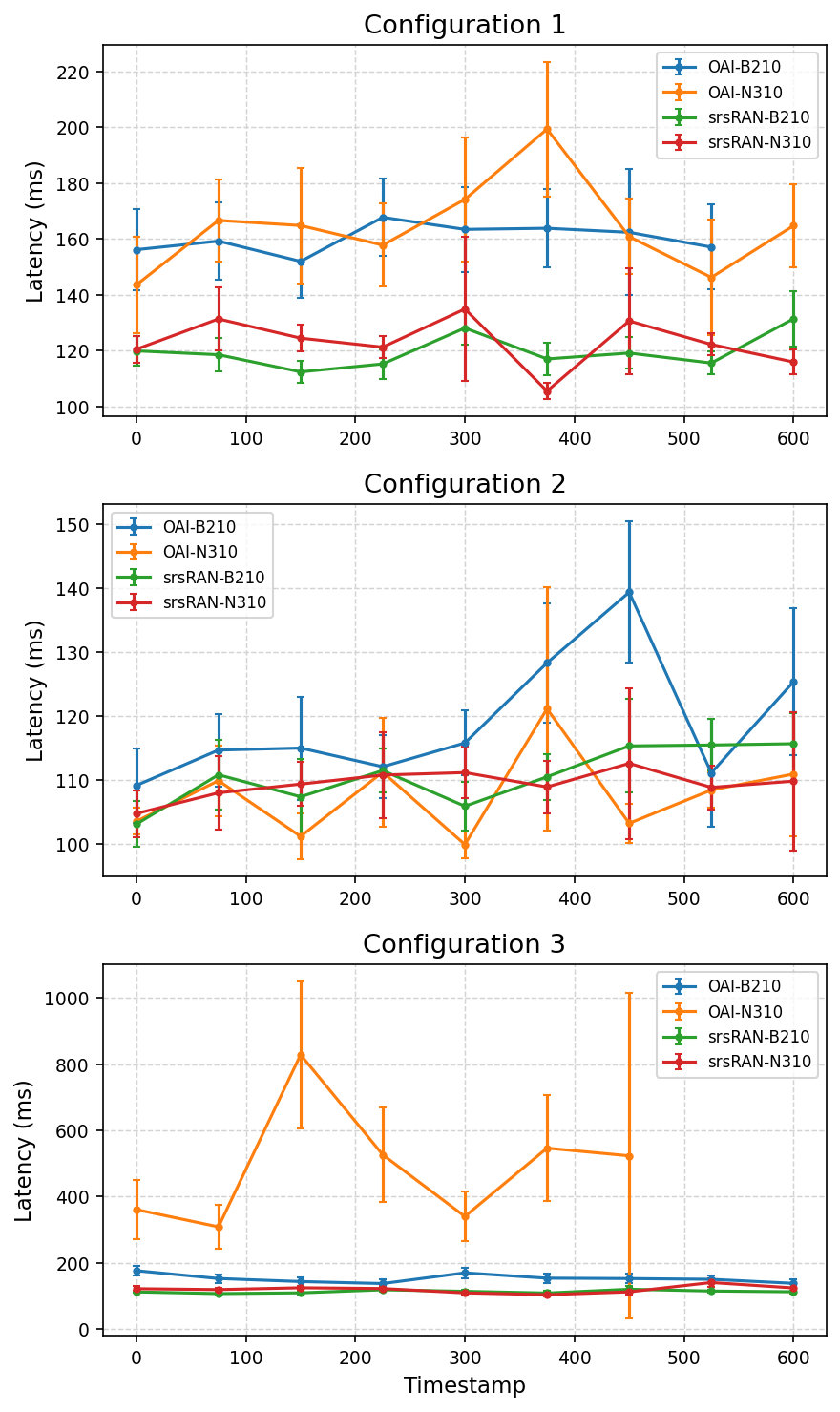}
    \caption{Latency comparison for LS across different configurations and 5G stacks.}
    \label{fig:ls}
\end{figure}

Live Streaming (LS) is a latency-sensitive application, and the objective of this flow is to evaluate network latency for each platform combination. To broadcast the live video, we used the open-source software Owncast\footnote{https://owncast.online/}, configured with a bitrate of 12~Mbps and a frame rate of 24~FPS. The LS server was running locally on a Lenovo ThinkPad E14 G2, which was connected to the 5G data plane. OBS Studio \cite{obsstudio} was used to stream the video to Owncast, which was responsible for forwarding it to the UEs. On the user side, a Kotlin-based Android application continuously requested the video using a media player (e.g., ExoPlayer) and collected and stored latency measurements in a CSV file. The test involved four UEs making requests simultaneously. The results are shown in Figure ~\ref{fig:ls}.

In C1, the OAI stack shows significantly worse performance than srsRAN, with 33.91\% higher latency, and even experiences connection loss with the B210. For srsRAN, the B210 achieved the best result, with a mean latency of about 120~ms, and with the N310 was only 2.5\% worse then with B210, indicating similar performance. In C2, the worst case was OAI with the B210, again, reaching a maximum latency of 320~ms and a mean of 120~ms. In contrast, with N310 provided one of the best performances, with a mean of 108~ms and 109~ms for OAI and srsRAN, respectively, indicating that the N310 is the most suitable SDR for this configuration. In C3, OAI with the N310 had the worst performance overall, with unstable behavior and a mean of 485.09~ms. By contrast, srsRAN showed stable behavior, though with slightly higher mean latencies 113~ms on B210 and 119~ms on N310 compared to other configurations. Overall, OAI generally shows higher latency, except for C2 with the N310, while srsRAN provides more stable results with consistently lower mean values.

\subsubsection{Cloud Gaming}

The second latency-sensitive application evaluated was Cloud Gaming (CG). In the experiments, a local server (Lenovo ThinkPad E14 G2) ran Sunshine\footnote{https://app.lizardbyte.dev/Sunshine/?lng=en}, integrated with Steam Remote Play\footnote{https://store.steampowered.com/remoteplay}, to stream the game Brawlhalla\footnote{https://store.steampowered.com/app/291550/Brawlhalla/}. On the client side, Moonlight\footnote{https://moonlight-stream.org/}, a streaming application based on the NVIDIA Gamestream protocol, was used to establish the connection. The client was configured with a frame rate of 60~FPS, a bitrate of 20~Mbps, and a resolution of 1080p. 

\begin{figure*}[ht!]
    \centering
    \includegraphics[width=0.85\linewidth]{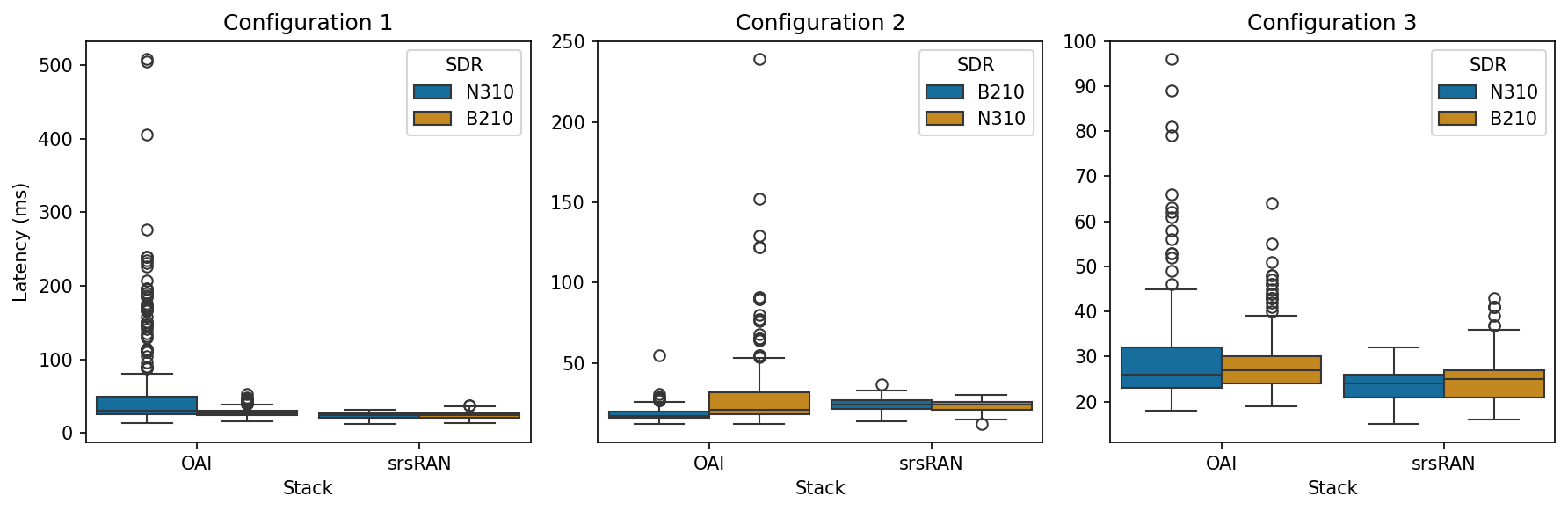}
    \caption{Latency comparison for CG across different configurations and 5G stacks.}
    \label{fig:cg_latency}
\end{figure*}

\begin{figure*}[ht!]
    \centering
    \includegraphics[width=0.85\linewidth]{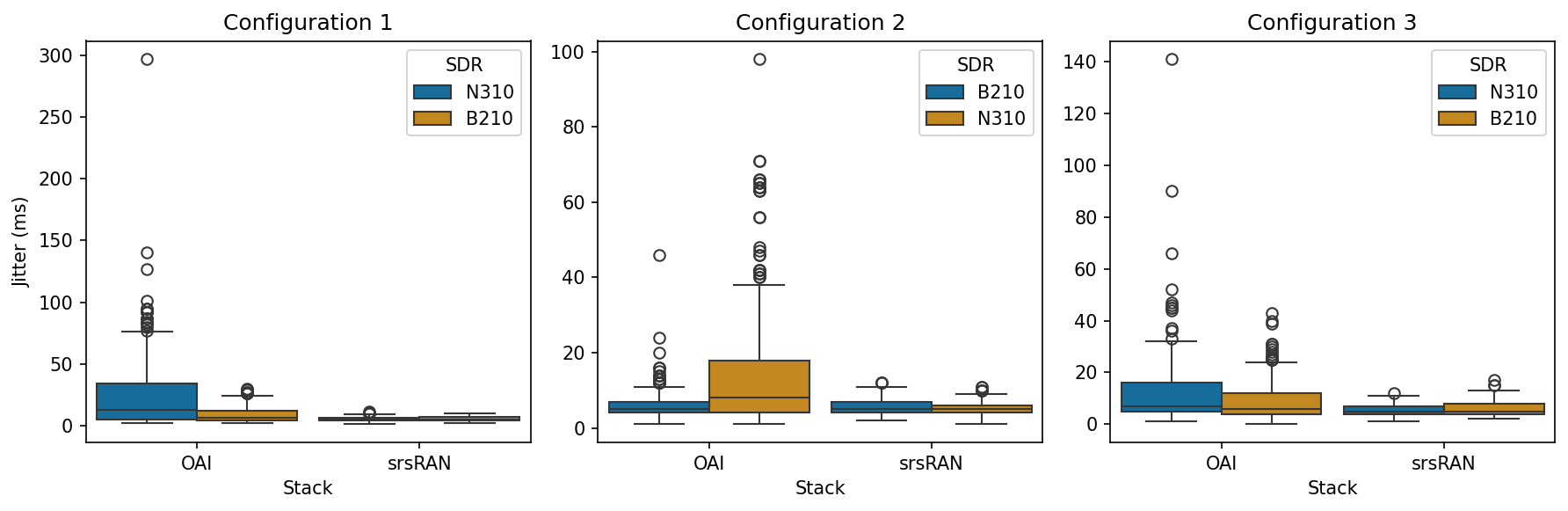}
    \caption{Jitter comparison for CG across different configurations and 5G stacks.}
    \label{fig:cg_jitter}
\end{figure*}

For the test, we used three UEs, and each played Brawlhalla simultaneously in offline mode, ensuring that latency was generated only in the 5G environment. Latency metrics were collected by enabling Moonlight’s on-screen display, recording the device screen, and parsing the metrics into a CSV file using a Python Code with the Gemini API. The results are presented in Figure~\ref{fig:cg_latency} and Figure~\ref{fig:cg_jitter}. Analyzing the srsRAN stack, all configurations have a similar behavior, with a mean latency of 24,06ms, with the best configuration in this stack being with C2 on SDR N310, with mean latency 23,54 ms and mean jitter 5,31ms. In opposition to srsRAN, OAI had a behavior unstable, with high jitter, having several outliers, represented by the boxplot, except for the configuration two on the SDR B210, which had the best performance, with mean latency 68,70\% better than the worse configuration (OAI with SDR N310). Even with high latency and high jitter in some cases, the frame dropped was close to zero.
\section{Conclusion}
\label{sec:conc}

This paper presented an overview of the leading open-source 5G RAN platforms, followed by a preliminary performance analysis of these stacks. The qualitative analysis highlighted the main features and supported capabilities of each platform. OAI and srsRAN share many similarities, with OAI distinguished by its support for 120~kHz SCS in FR2, while srsRAN is notable for its well-structured documentation. The quantitative analysis showed that OAI achieved the best results for the RRC Setup procedure, however, the overall performance of both platforms were similar. For the data-plane evaluation, regarding TT for downlink srsRAN achieved results closest to the theoretical values, reaching up to 70\% of TT across all SDRs and configurations. In uplink, srsRAN also performed best, achieving about 30\% of TT in all scenarios. For real application workloads, srsRAN consistently delivered the best results across all evaluated cases. Future work will focus on testing higher bandwidths and evaluating the computational costs of deployment.

\section*{Acknowledgment}

This work was supported by the Motorola Mobility, National Council for Scientific and Technological Development (CNPq) - Research Productivity Fellowship (Grant No. 313083/2023-1) and Pernambuco Research Foundation (FACEPE) (Grant No. IBPG-0130-1.03/23).

\end{document}